\documentclass[letterpaper]{article} 
\usepackage{aaai2026}  
\usepackage{times}  
\usepackage{helvet}  
\usepackage{courier}  
\usepackage[hyphens]{url}  
\usepackage{graphicx} 
\usepackage{natbib}  
\usepackage{caption} 
\usepackage[ruled, vlined, linesnumbered]{algorithm2e}

\title{Cost-Free Neutrality for the River Method}

\title{Cost-Free Neutrality for the River Method}
\author {
    Michelle Döring\textsuperscript{\rm 1},
    Jannes Malanowski\textsuperscript{\rm 1},
    Stefan Neubert\textsuperscript{\rm 1}
}

\affiliations {
    \textsuperscript{\rm 1} Hasso Plattner Institute, University of Potsdam, Germany\\
    michelle.doering@hpi.de, jannes.malanowski@student.hpi.de, stefan.neubert@hpi.de
}

\newif\iflong
\newif\ifshort
\longtrue

\iflong
\else
\shorttrue
\fi

\usepackage[colorinlistoftodos]{todonotes}

\usepackage{xspace}
\usepackage{amsmath}
\usepackage{mathtools}
\usepackage{amssymb}

\usepackage{amsthm}

\newtheorem{theorem}{Theorem}[section]
\newtheorem{definition}[theorem]{Definition}
\newtheorem{lemma}[theorem]{Lemma}
\newtheorem*{lemma*}{Lemma}

\newtheorem*{observation*}{Observation}

\usepackage{thm-restate}

\newtheoremstyle{claimstyle}{\topsep}{\topsep}{}{0pt}{\sffamily}{. }{5pt plus 1pt minus 1pt}%
  {$\vartriangleright$ \thmname{#1}\thmnumber{ #2}\thmnote{ (#3)}}
\theoremstyle{claimstyle}
\newtheorem{claim}[theorem]{Claim}
\newtheorem*{claim*}{Claim}

\usepackage{tabularray}
\usepackage{cleveref} 
\crefname{claim}{claim}{claims}
\Crefname{claim}{Claim}{Claims}
\usepackage{csquotes}

\newcommand{\defstyle}[1]{\emph{#1}}
\newcommand{\ie}{i.\,e., }
\newcommand{\abs}[1]{\left\lvert #1 \right\rvert}
\newcommand{\sset}[1]{\{#1\}}
\newcommand{\set}[2]{\{\, #1 \mid #2 \,\}}
\newcommand{\NP}{\ensuremath{\mathtt{NP}}\xspace}

\newcommand{\Pref}{\ensuremath{\mathcal{P}}\xspace}
\newcommand{\pref}{\ensuremath{\succ}\xspace}

\newcommand{\margin}[1]{\ensuremath{m(#1)}\xspace}
\newcommand{\marginExt}[2]{\ensuremath{m_{#2}(#1)}\xspace}
\newcommand{\mgg}{\ensuremath{\mathcal{M}}\xspace}

\newcommand{\mggExtend}[1]{\ensuremath{\mathcal{M}^{#1}}}
\newcommand{\mggRiver}{\ensuremath{\mggExtend{\RV}}\xspace}

\newcommand{\strength}{\operatorname{strength}}
\newcommand{\Path}[2]{\ensuremath{P_{#1#2}}\xspace}

\newcommand{\lino}{\ensuremath{\tau}\xspace}
\newcommand*{\allDescOrders}{\ensuremath{\mathbb{L}}}

\newcommand{\RV}{\ensuremath{\mathsf{RV}}\xspace}
\newcommand{\RVput}{\ensuremath{\operatorname{\mathsf{RV\mbox{-}PUT}}}\xspace}
\newcommand{\RPput}{\ensuremath{\operatorname{\mathsf{RP\mbox{-}PUT}}}\xspace}

\newcommand{\FUN}{\ensuremath{\mathsf{FUN}}\xspace}
\newcommand{\FUNsym}{\ensuremath{{\mgg^{\FUN}}}\xspace}
    
\newcommand{\CertTreeAlgo}{\textsc{DirectedMaxPrim}\xspace}
\newcommand{\CertTreeSym}{\ensuremath{T^a}\xspace}
\newcommand{\FUNlinoSym}{\ensuremath{\lino^a}\xspace}
\newcommand{\FUNlinoSymNAME}{certificate tiebreaker\xspace}
\newcommand{\Efix}{\textnormal{\textsf{Fix}}\xspace}
\newcommand{\EBC}{\textsf{BC}\xspace}
\newcommand{\ECC}{\textsf{CC}\xspace}
\newcommand{\ECBC}{\textsf{CBC}\xspace}

\begin{document}

\maketitle

\begin{abstract}
    Recently, the River Method was introduced as novel refinement of the Split Cycle voting rule.
    The decision-making process of River is closely related to the well established Ranked Pairs Method.
    Both methods consider a margin graph computed from the voters' preferences and eliminate majority cycles in that graph to choose a winner.
    As ties can occur in the margin graph, a tiebreaker is required along with the preferences.
    While such a tiebreaker makes the computation efficient, it compromises the fundamental property of neutrality: the voting rule should not favor alternatives in advance.
    
    One way to reintroduce neutrality is to use Parallel-Universe Tiebreaking (PUT), where each alternative is a winner if it wins according to any possible tiebreaker.
    Unfortunately, computing the winners selected by Ranked Pairs with PUT is \NP-complete.
    Given the similarity of River to Ranked Pairs, one might expect River to suffer from the same complexity.
    
    Surprisingly, we show the opposite:
    We present a polynomial-time algorithm for computing River winners with PUT, highlighting significant structural advantages of River over Ranked Pairs.
    Our Fused-Universe (FUN) algorithm simulates River for every possible tiebreaking in one pass.
    From the resulting FUN diagram one can then directly read off both the set of winners and, for each winner, a certificate that explains how this alternative dominates the others.
\end{abstract}


\section{Introduction and Related Work}

A common interest in theoretical computer science, economics, and political science is the design and analysis of \textit{social choice functions}-mechanisms that aggregate individual preferences to make a collective decision.
Clearly, such decision-making is an integral part of democratic processes.
%
    Beyond political contexts, social choice functions have become increasingly relevant in artificial intelligence, particularly for aligning AI systems with diverse human preferences \cite{furnkranz2003pairwise,kopf2024openassistant}.
    As language models and other AI systems interact with feedback from a wide range of users, aggregating that feedback `socially acceptably' becomes a key challenge. Recent work argues that social choice theory should guide AI alignment in addressing this diversity \cite{conitzer_social_2024}.
    
    There is extensive research on properties of social choice functions, their trade-offs, and the often high complexity of computing them (see \cite{brandt_handbook_2016} for an overview).

A core task of computational social choice is to design rules that satisfy reasonable fairness criteria while also being computationally tractable.
In this work, we show how to efficiently compute the winners of the recently suggested \textit{River Method} \cite{doring_river_2025}.

\iflong
\subsection{Margin-Based Social Choice Functions}\fi
    River is part of a family of \textit{margin-based social choice functions} -- methods that decide an election based on pairwise comparisons between alternatives and the \textit{margin} of victory in each such comparison.
    A common approach to social choice is to select an alternative that wins all pairwise comparisons; the so-called Condorcet winner. 
    However, there are cases where there is no such alternative.
    Margin-based rules resolve this by not just considering who wins in a pairwise comparison, but also by how much.

    The \textit{margin} of $x$ over $y$ is the difference between the number of voters who prefer $x$ over $y$ and those who prefer $y$ over $x$.
    This information is represented as a \emph{margin graph} -- a complete, weighted, antisymmetric, directed graph over the set of alternatives, where each edge $(x, y)$ is weighted by the margin between $x$ and $y$ \cite{brandt_handbook_2016}.
    If there is no Condorcet winner, the margin graph contains cycles of pairwise victories, such as alternatives $x,y,z$ where $x$ wins against $y$, which wins against $z$, which wins against $x$.

    One way of resolving such cycles is to focus on the notion of \emph{immunity to majority complaints}:
    an alternative $x$ is \emph{immune} if for every alternative $z$ that beats $x$ in direct comparison, there exists a path of at least as strong pairwise victories leading from $x$ to $z$.
    This path does not only serve to define a winner, but also to defend the choice of $x$ against claims such as \enquote{a majority prefers $z$ over $x$, thus $x$ should not win}.

    \iflong
        In every election there exists at least one immune alternative \cite{holliday_split_2023}. 
        Among the numerous social choice functions selecting immune alternatives, we want to highlight five that are particularly notable for their distinct methods of resolving majority cycles:
        \begin{itemize}
            \item \textbf{Split Cycle} \cite{holliday_split_2023} aims to identify a consistent winner set by deleting the weakest link in every cycle in the margin graph.
                As Split Cycle selects \textit{all} immune alternatives as winners, the following methods can be seen as refinements.
            \item \textbf{Ranked Pairs} \cite{tideman_independence_1987} incrementally locks in pairwise victories, prioritizing those with the largest margins that do not create cycles.
            \item \textbf{Stable Voting} \cite{holliday_stable_2023} recursively refines the winner set, ensuring stability under the deletion of alternatives.
            \item \textbf{Beat Path} \cite{schulze_new_2011,schulze_schulze_2025} selects winners based on the strongest paths through the defeat graph.
            \item \textbf{River} \cite{doring_river_2025} builds on Ranked Pairs by providing a simpler approach to refine the majority cycle elimination process.
        \end{itemize}
        For an overview of the definitions and properties of these methods, we refer to \cite{holliday_split_2023, doring_river_2025}.
        
        Among these margin-based choice functions, Ranked Pairs, Stable Voting, and River share a key feature: they process the edges of the margin graph in decreasing order of the margin.
        When multiple edges have the same margin, these methods require a tiebreaking rule (a total descending linear ordering of the margin edges) as additional input.
        
        Unfortunately, such a tiebreaker usually compromises the property of \textit{neutrality}, which requires that no alternative is favored a priori. See \cite{brill_price_2012,freeman_general_2015,wang_practical_2019,zavist_complete_1989} for a discussion of possible tiebreaking rules and their implications.
        To reintroduce neutrality, one can compute the winning set using the \textit{Parallel-Universe Tiebreaking} (PUT), which considers all possible tiebreaking orders and returns the union of winners.
        Unfortunately, deciding whether an alternative is a winner under PUT for Ranked Pairs is known to be \NP-complete \cite{brill_price_2012}.
    \fi

\subsection{Our Contribution}

We study the computational complexity of applying Parallel-Universe Tiebreaking (PUT) to River.
Although River closely resembles Ranked Pairs, we show that the winners under PUT can be computed in polynomial time.

For this, we exploit the more restricted rule set of River:
like Ranked Pairs, it processes the edges of the margin graph in order of decreasing margin and discards those that would create a cycle; River additionally enforces that each alternative has at most one incoming edge.
As a result, River always produces a directed tree rooted at the winner, whereas Ranked Pairs only guarantees to return an acyclic subgraph.

This difference enables us to simultaneously simulate River for all possible tiebreakers \textit{in one pass}.
Our \textit{Fused-Universe} (\FUN) algorithm to determine all PUT winners for River can be summarized as follows:

\begin{enumerate}
    \item Compute the Fused-Universe diagram: a subgraph of the margin graph, which includes all margin edges that appear in the River diagram of at least one universe. 
    \item During that same process, track and update for each alternative whether it remains undefeated in all universes, is defeated in some, or is defeated in all universes.
    \item Derive the set of winners from these vertex states.
\end{enumerate}

After \iflong introducing some notation and \fi formally defining River in \Cref{sec:preliminaries}, we describe our algorithm in detail in \Cref{sec:Algorithm}, where we also explain how it can be computed in polynomial time.
In \Cref{sec:correctness} we prove that the algorithm correctly determines the set of \RVput winners.
We first show that every alternative not selected by our algorithm is defeated in every universe.
For the other direction, we construct, for each selected winner, a tiebreaker under which it wins in River.
This tiebreaker not only proves correctness but also serves as a constructive certificate that the alternative is \textit{immune}.
We conclude in \Cref{sec:conclusion} with open questions and suggestions for future work on River and Parallel-Universe Tiebreaking.

\ifshort
Proofs of statements marked with $\star$ are in the full version.
\fi

\section{Preliminaries} \label{sec:preliminaries}

We consider $n\ge 1$ \defstyle{voters} expressing preferences over a set $A$ of \defstyle{alternatives}.
The preferences of each voter $i$ are represented by a linear order ${\pref_i}$ over $A$,  and the \defstyle{(preference) profile} $\Pref={(\pref_1,\dots,\pref_n)}$ collects these preferences for all voters.
Let $[n] = \{1,\dots,n\}$.
The \defstyle{margin} of alternative $x$ over $y$ according to \Pref is
$\marginExt{x,y}{\Pref} = \abs{\set{i \in [n]}{x \pref_i y}} - \abs{\set{i \in [n]}{y \pref_i x}}$. \iflong

\fi The \emph{margin graph} $\mgg(\Pref)$ of~$\Pref$ is a weighted directed graph with vertex set $A$ and edge set $\set{(x,y)\in A\times A}{\marginExt{x,y}{\Pref} \geq 0}$, where each edge $(x,y)$ is assigned weight $\marginExt{x,y}{\Pref} \geq 0$.
In any graph with vertex set $A$, we say that vertex $y$ is \emph{dominated} by the edge $(x,y)$ or by vertex $x$, if $y$ has an incoming edge from $x$.
We omit the preference profile \Pref when it is clear from the context\iflong;
for example, we write $\margin{x,y}$ for a margin function and $\mgg$ for a margin graph\fi. 

A \defstyle{(majority) path} from $x$ to $y$ in $\mgg$ is a sequence $\Path{x}{y} = (x = p_1, \ldots, p_\ell = y)$ of distinct alternatives such that $\margin{p_i, p_{i+1}} > 0$ for all $i \in [\ell-1]$.
The \defstyle{strength} of such a path is the value of its lowest margin, \ie 
$\strength(\Path{x}{y}) = \min\set{\margin{p_i, p_{i+1}}}{i \in [\ell-1]}$.
Analogously, \defstyle{majority cycles} in $\mgg$ are closed paths ($x_\ell = x_1$) such that $\margin{x_i,x_{i+1}} > 0$ for all consecutive pairs, and their strength is defined as the lowest margin on the cycle.

\subparagraph{Winners and Ties}
\iflong A \defstyle{Condorcet winner} $x$ is an alternative which defeats all other alternatives in a pairwise comparison, \ie $\margin{x,y} > 0$ for all $y \in A \setminus \sset{x}$.
Such an alternative does not always exist.
\fi
In the absence of a Condorcet winner, every alternative faces at least one majority defeat.
Some alternatives can be justified as winners despite these defeats by the existence of equally strong majority paths leading from the alternative to each alternative defeating it.
\iflong This notion is called immunity \cite{holliday_split_2023}:
\fi
An alternative $x$ is \defstyle{immune (against majority complaints)} if, for every alternative $y$ with $\margin{y,x}>0$, there is a majority path $\Path{x}{y}$ with $\strength(\Path{x}{y})\geq \margin{y,x}$.

A \textit{social choice function} maps a preference profile to a non-empty set of \textit{winning} alternatives.
For margin graphs where at least two edges have the same margin, Stable Voting, Ranked Pairs, and River require a tiebreaker to compute their winning set.
A \textit{tiebreaker} is a \textit{descending linear ordering} $\lino = (e_1, \dots, e_{\lvert E\rvert})$ of the margin edges $E$ by decreasing margin, \ie $\margin{e_i} \geq \margin{e_j}$ for all $1 \leq i < j \leq \lvert E\rvert$ \cite{zavist_complete_1989,brill_price_2012}.
We refer to the set of all descending linear orderings by $\allDescOrders$.

\paragraph{The River Method}
River \cite{doring_river_2025} is a social choice function which eliminates majority cycles in the margin graph by greedily constructing a subtree of \mgg based on a fixed tiebreaker.
Formally, given a preference profile \Pref and a tiebreaker $\lino \in \allDescOrders$, let $\mgg = (A, E)$ denote the corresponding margin graph.
River constructs the \textit{River diagram} $\mggRiver(\Pref, \lino)$, and selects a unique winner $\RV(\Pref, \lino)$ as follows:

\begin{enumerate}
    \item Initialize $\mggRiver(\Pref, \lino)$ as $(A, \emptyset)$.
    \item Iterate over the edges in order of $\tau$ and add $(x,y)$ to the diagram unless it is rejected by one of two conditions:
        \begin{description}
            \item[branching condition:] 
            $y$ already has an incoming edge; 
            \item[cycle condition:] adding the edge would create a cycle.
        \end{description}
    \item Return the unique alternative without incoming edges in $\mggRiver(\Pref,\lino)$ as winner $\RV(\Pref, \lino)$.
\end{enumerate}

\paragraph{Parallel-Universe Tiebreaking}
Instead of fixing one tiebreaker, Parallel-Universe Tiebreaking (PUT) selects every winner according to any possible tiebreaker.
Formally,
\[\RVput(\Pref) = \bigcup_{\lino\in\allDescOrders} \RV(\Pref,\lino).\]

\section{Fusing Parallel Universes} \label{sec:Algorithm}
    
    \iflong 
    The \FUN algorithm efficiently simulates the River procedure across all universes simultaneously.
    To do so, it exploits River's unique branching condition: every alternative has at most one incoming edge in any River diagram.
    This allows us to reason about which edges might be added or rejected across all possible universes.
    \fi

    Consider \iflong the margin graph in \fi \Cref{fig:edge-states}. The edges $(x,y)$, $(z,y)$, and $(d,y)$, with margin~$10$, are processed first by River. Depending on the tiebreaker, one is added, while the other is rejected by the branching condition.
    As a result, any edge to $y$ with margin lower than 10 
    is rejected by the branching condition regardless of the tiebreaker.
    This allows us to reason about potential winners: once all margin-10 edges are processed, it is already certain that $y$ cannot be a River winner in any universe. Indeed, one of $(x,y)$, $(z,y)$, or $(d,y)$ will appear in every River diagram, thereby dominating $y$. Consequently, $y$ cannot be an \RVput winner.

\definecolor{Mercury}{rgb}{0.917,0.917,0.917}
\definecolor{Alto}{rgb}{0.854,0.854,0.854}

\begin{table*}
    \centering
    \begin{tblr}{
        colspec={l Q[l] Q[l] Q[l]}, 
        cells = {Mercury},
        row{even} = {Alto},
        vline{2-4} = {-}{white},
        hline{2} = {-}{},
        column{1} = {0.04\textwidth}, 
        column{2} = {0.125\textwidth}, 
        column{3} = {0.4\textwidth}, 
        column{4} = {0.34\textwidth}, 
    }
        \textit{short}    & \textit{edge state} & \textit{interpretation}  & \textit{implication} \\
        \Efix    & fix                           & added in every universe & 
        in every universe, $y$ is dominated by $(x,y)$  \\
        \EBC   & branching choice              & added in some but not all universes; there is $(z,y)$ with $\margin{z,y}=k$ which could be added instead & 
        in every universe, $y$ is dominated by an edge with margin $k$ \\
        \ECC   & cycle choice                  & may be added in some but not all universes; there exists an $y$-$x$-path which could be added instead &
        there is a universe where $y$ is not dominated by $(x,y)$    \\
        \ECBC & cycle branching choice        & 
        added in some but not all universes; there exists $(z,y)$ with $\margin{z,y}>k$ which is \ECC &  
        in every universe, $y$ is dominated by an edge with margin at least $k$ \\
    \end{tblr}
    \caption{Overview of the possible edge states assigned to an edge $(x,y)$ with margin $k=\margin{x,y}$.}
    \label{tab:FUN_alg-edge-states}
\end{table*}
\vspace{0.4em}\textbf{The \FUN algorithm.}\quad
Our algorithm to compute the \RVput winners has two main steps.

In the first step, it computes the \textit{Fused-Universe} (\FUN) diagram.
The algorithm takes as input the margin graph and processes its edges in order of decreasing margin; edges with the same margin are considered in arbitrary order.
For each edge $(x,y)$ with margin $k$ the algorithm checks whether the edge is clearly rejected from every possible universe because of River's branching or cycle condition.
If the algorithm has decided that $(x,y)$ cannot be clearly rejected from every universe, it adds the edge to the \FUN diagram and proceeds to compute the conditions under which this edge may appear in some universe.
These conditions are represented by four possible \textit{edge states}:
fix, branching choice, cycle choice, or cycle-branching choice.
Each alternative -- represented in the diagram as a vertex -- also receives a \textit{vertex state}: not dominated, fixedly dominated, or cycle dominated.

In the second step, the \FUN algorithm returns the set of winning alternatives, which is the set of vertices which are not dominated or cycle dominated in the \FUN diagram.

We introduce the edge and vertex states in \Cref{subsec:edge-states}, followed by a detailed description of their computation in \Cref{subsec:compute-diagram}.
Afterwards, we explain in \Cref{subsec:algorithm-runtime} how the algorithm can be implemented to run in polynomial time.

\ifshort 
    \begin{figure}
        \centering
        \includegraphics[width=1.0\columnwidth]{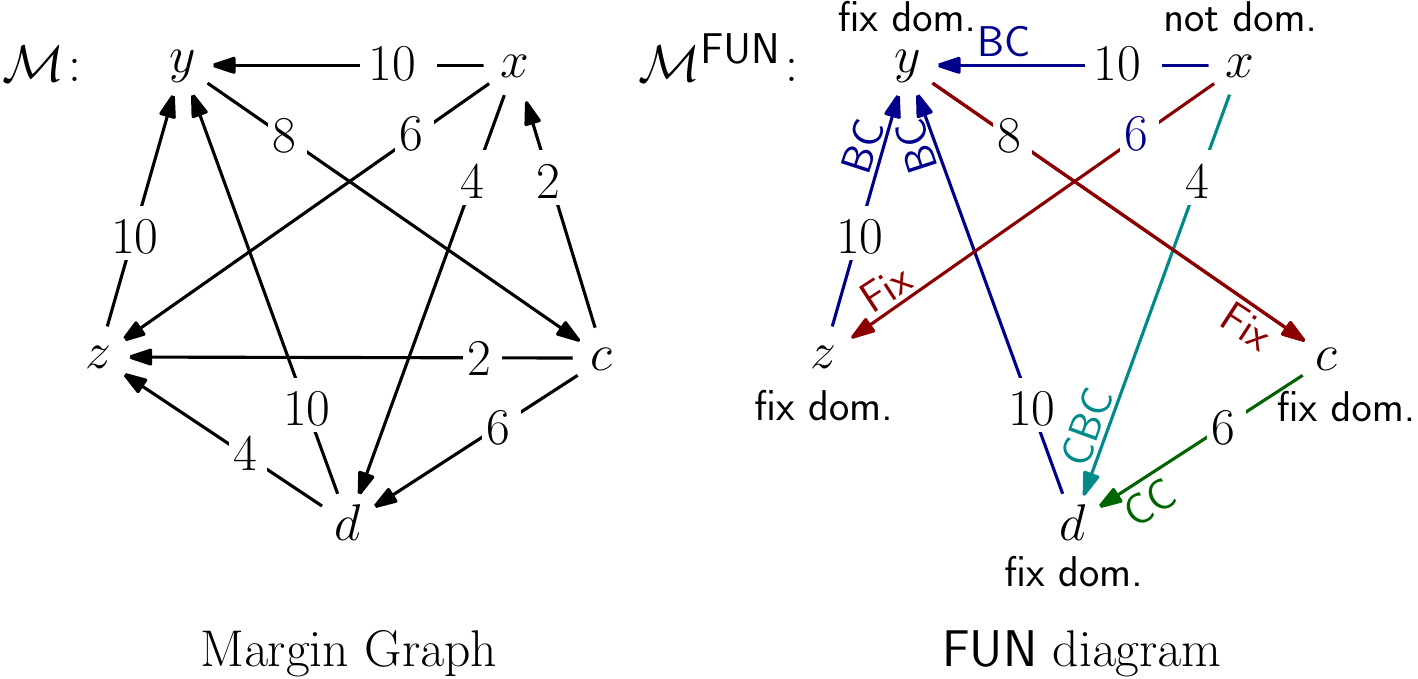}
        \captionof{figure}{Election with margin graph $\mgg$, \FUN diagram \FUNsym with edge/vertex states and \RVput winner $x$.}
        \label{fig:edge-states}
    \end{figure}
\fi
\subsection{Edge and Vertex States}\label{subsec:edge-states}
\paragraph{Edge States}
    \iflong If run with \fi a fixed tiebreaker, River either rejects or adds an edge to the River diagram.
    Our algorithm is more nuanced about the edge states, as it considers simultaneously whether edges are rejected or added in any possible universe.
    An edge in the \FUN diagram receives one of four edge states\iflong -- fix, branching choice, cycle choice, and cycle branching choice\fi.
    Their interpretations, implications, and abbreviations are summarized in \Cref{tab:FUN_alg-edge-states} and illustrated in \Cref{fig:edge-states}.
    Let $(x,y)$ with margin $k$ be the current edge.
    
    \ifshort
        A \textit{fix} edge appears in every universe.
        An edge is marked \Efix if we are certain no other incoming edge to $y$ has margin $\geq k$, and no path from $y$ to $x$ has strength $\geq k$.
        A simple example is a unique edge with maximum margin: since it is processed first, neither River's branching nor cycle condition reject it and the edge is added to every River diagram.
    \else
        A \textit{fix} edge appears in every universe.
        An edge is marked \Efix if we are certain that no other incoming edge to $y$ of margin $\geq k$ and no path from $y$ to $x$ has strength $\geq k$.
        A simple example for a \Efix edge is a unique edge with maximum margin: since that edge is considered as the very first edge in every universe, neither River's branching nor cycle condition reject it and the edge is added to every River diagram.
    \fi
    
    A \textit{branching choice} edge may appear in some universe but cannot appear in all.
    That is because there exists at least one other incoming \enquote{partner} edge $(z,y)$ with margin $k$ which also could be added.
    Which of the two edges is added to the River diagram depends on the order in the tiebreaker $\tau$: if $(x,y)$ is before $(z,y)$ in $\tau$, then $(x,y)$ is added and $(z,y)$ must be rejected by the branching condition.
    Vice versa, if $(z,y)$ is before $(x,y)$ in $\tau$, then $(z,y)$ is added and $(x,y)$ rejected.
    Both edges are marked as a branching choice \EBC.
    
    \iflong
        A \textit{cycle choice} edge may appear in some but not all universes. 
        Assume there is a path from $d$ to $c$ containing at least one \EBC edge $(d,y)$, and let $(z,y)$ be the partner \EBC that is not on the path.
        If the tiebreaker puts $(z,y)$ first, $(d,y)$ is rejected by the branching condition.
        Thus, the addition of $(c,d)$ does not close a cycle and the edge is added.
        If $(d,y)$ is put first instead, the addition of $(c,d)$ would close a cycle, and it would be rejected by the cycle condition.
        Another simple example would be a directed cycle $(a,b,c)$ containing only edges of margin $k$.
        Depending on the order in the tiebreaker, two of the three edges will be added to the River diagram while the third is rejected by the cycle condition.
    \fi
    
    \iflong A \textit{cycle branching choice} edge is a branching choice which is dependent on the cycle choice of another edge.
    Assume $(c,d)$ is a cycle choice edge because there is a path from $d$ to $c$ that includes a branching choice edge $(d,y)$ with partner $(z,y)$ (not on the path).
    In that case, the addition of $(x,d)$ depends on the choices made on the cycle and might lead to a rejection by the branching condition:
    if $(z,y)$ is processed first and added, then $(d,y)$ is rejected, $(c,d)$ does not close a cycle and is added; consequently, $(x,d)$ must be rejected by the branching condition.
    If $(d,y)$ is processed and added first instead, then $(c,d)$ is rejected by the cycle condition and $(x,d)$ is added.
    \fi

\paragraph{Vertex States}
\begin{table}
    \centering
    \begin{tblr}{
        colspec={l Q[l]}, 
        cells = {Mercury},
        row{even} = {Alto},
        vline{2} = {-}{white},
        hline{2} = {-}{},
        column{1} = {0.25\columnwidth}, 
        column{2} = {0.65\columnwidth}, 
    }
        \textit{vertex state} & \textit{property in $\FUNsym$} \\
        not dominated         & no incoming edge    \\
        fixedly dominated     & {either exactly one incoming edge with state \Efix, or at least one incoming edge with state \EBC or \ECBC} \\
        cycle dominated       & at least one incoming edge and all incoming edges have state \ECC
    \end{tblr}
    \caption{Overview of the possible vertex states.}
    \label{tab:Fun_alg-vertex-states}
\end{table}
\iflong
\begin{figure}
    \centering
    \includegraphics[width=1.0\columnwidth]{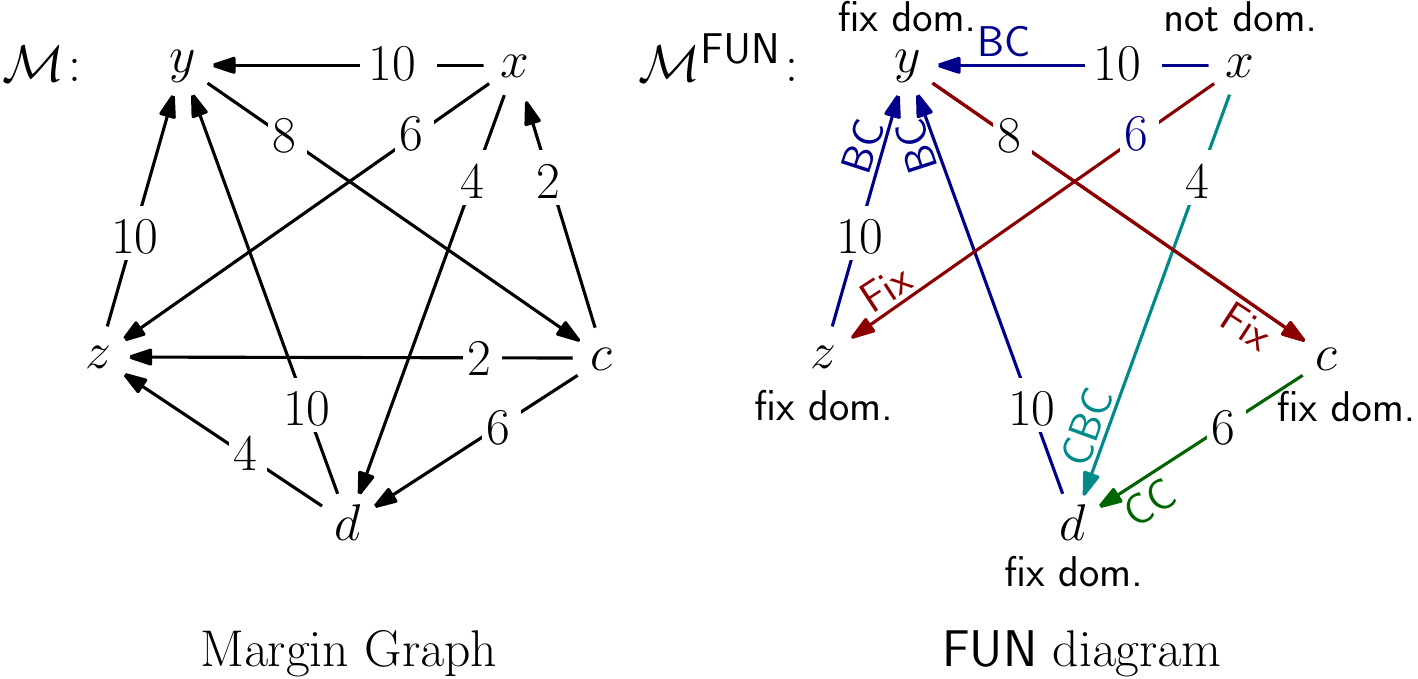}
    \captionof{figure}{Election with margin graph $\mgg$, \FUN diagram \FUNsym with edge/vertex states and \RVput winner $x$.}
    \label{fig:edge-states}
\end{figure}
\fi

\iflong 
    If run with a fixed tiebreaker, River selects a winner after all edges have been processed -- the unique alternative with no incoming edge in the resulting River diagram.
    Our algorithm generalizes this by assigning \textit{vertex states} during execution, capturing structural information about which alternatives can or cannot be winners across all universes.
\fi 

These vertex states reflect what is known about a vertex $y$'s incoming edges in \FUNsym.
Edges labeled \Efix, \EBC, or \ECBC with margin $k$ indicate that $y$ is dominated by an edge of margin~$\geq k$ in \emph{every} universe.
In contrast, \ECC edges do not guarantee such domination, as they may be rejected depending on the cycle structure and tiebreaker.
This distinction allows us to assign each vertex one of three possible states, as summarized in \Cref{tab:Fun_alg-vertex-states} and illustrated in \Cref{fig:edge-states}.

\subsection{Computing the Fused-Universe Diagram}
\label{subsec:compute-diagram}

Let us now go into the details on how to compute the introduced edge and vertex states for the \FUN diagram.
Given a margin graph $\mgg$, the algorithm processes the edges in descending order of their margin, breaking ties arbitrarily.
\Cref{alg:RVPUT} shows how each edge $(x, y)$ with margin $\margin{x,y} = k$ is then processed in two phases:

\newcommand{\GrayComment}[1]{\textcolor{gray}{#1}}
\SetCommentSty{GrayComment}

\begin{algorithm*}[h]
\DontPrintSemicolon
\KwIn{margin graph $\mathcal{M(\Pref)} = (A, E)$ of a preference profile \Pref with margins $m$}
\KwOut{\FUN diagram $\FUNsym$}

\BlankLine 
\BlankLine 

$\FUNsym \gets (A, \emptyset)$\label{line:init}\;
\BlankLine
\ForEach{$(x,y) \in E$ in order of decreasing margin\label{line:loop}}{
    $k \gets \margin{x,y}$\;

    \iflong\BlankLine
    \tcp{branching reject check}\fi
    \lIf{$y$ is fixedly dominated and $\exists (z,y) \in \FUNsym \colon \margin{z,y} > k$ \label{line:begin branching reject check}}{
        \KwSty{reject} $(x,y)$ and \KwSty{continue} \label{line:end branching reject check}
    }

    \iflong\BlankLine
    \tcp{cycle reject check}\fi
    \If{there is a path from $y$ to $x$ of strength $>k$ in \FUNsym \label{line:begin cycle reject check}}{
        compute the set of vertices $U$ that have a path to $x$ of strength $>k$ in \FUNsym \label{line:compute U}\;
        \If{for all $u \in U$ there exists a path $P \in \FUNsym$ from $y$ to $u$ of strength $> k$, such that for every \ECC edge $(c,d) \in P$, all paths from $d$ to $c$ in \FUNsym of strength $\geq \margin{c,d}$ include vertex $y$\label{line:check all u in U}}{
            \KwSty{reject} $(x,y)$ and \KwSty{continue} \label{line:end cycle reject check}\;
        }
    }

    \iflong\BlankLine
    \tcp{$(x,y)$ will be added to $\FUNsym$ -- compute tentative state of $(x,y)$}\fi
    \lIf{$y$ is not dominated \label{line:if not dominated} }{
        $state(x,y) \gets \Efix$ \label{line:assign state fix}
    }\lElseIf{$y$ is cycle dominated and $\exists (z,y) \in \FUNsym \colon \margin{z,y} > k$\label{line:if cycle dominated}}{
        $state(x,y) \gets \ECBC$ \label{line:assign state CyBrCh}
    }\lElseIf{$y$ is cycle dominated and $\forall (z,y) \in \FUNsym \colon \margin{z,y} = k$}{
        $state(x,y) \gets \EBC$
    }\Else(\label{line:if fixedly dominated}\iflong\tcp*[h]{$y$ is fixedly dominated by edges of margin $k$}\fi){
        $state(x,y) \gets \EBC$ \label{line:assign state BrCh}\;
        update incoming \Efix\ edge of $y$ to \EBC \label{line:update BrCh edges}\;
    }
    \textbf{add} $(x,y)$ to \FUNsym \label{line:add edge}\;

    \iflong\BlankLine
    \tcp{cycle update check}\fi
    compute the set of edges $C\subseteq \FUNsym$ that form a cycle with $(x,y)$ \label{line:begin CyCh update}\;
    \If{$C\neq\emptyset$}{
        $state(x,y) \gets \ECC$\;
        \lForEach{$e \in C$ with $\margin{e}=k$}{
            $state(e) \gets \ECC$ \label{line:end CyCh update}
        }
    }
}
\Return \FUNsym\;
\caption{Computing the fused universe diagram.}
\label{alg:RVPUT}
\end{algorithm*}


\iflong
\subparagraph{1. Rejection in Every Universe (lines \ref{line:begin branching reject check} to \ref{line:end cycle reject check}).}
\fi
The algorithm first decides whether the edge $(x,y)$ must be rejected from \FUNsym in every tiebreaker universe.
This decision is based on River’s branching and cycle conditions:
\begin{itemize}
    \item Branching rejection:
        If vertex $y$ is \textit{fixedly dominated} by some edge with margin~$>k$, then $(x,y)$ is rejected by River's branching condition in every universe.
    \item Cycle rejection:
        If adding $(x,y)$ forms a cycle with a path of strength~$> k$ in every universe, then in each of these universes the edge is rejected by River's cycle condition.
        This is checked by the algorithm in lines~\ref{line:begin cycle reject check} -- \ref{line:end cycle reject check}.
            
        The condition in line~\ref{line:begin cycle reject check} can be computed with a breadth-first search (BFS) started on $y$ that only considers edges with margin $> k$.
        Similarly, the set $U$ in line~\ref{line:compute U} can be computed with a reverse BFS started on $x$ that only considers edges with margin~$> k$.
        To check the condition in line~\ref{line:check all u in U}, we do the following.
        For every \ECC edge $(c,d)$ in \FUNsym, run a BFS started on $d$ that only considers edges with margin~$\geq \margin{c,d}$ that are not incident to~$y$.
        Let $E'$ be the set of such edges $(c,d)$ where this BFS reaches $c$ -- meaning there is a path from $d$ to $c$ in \FUNsym of strength~$\geq \margin{c,d}$ that does not include vertex $y$.
        Now the condition in line~\ref{line:check all u in U} can be tested by comparing~$U$ with the vertices reached by a BFS started on $y$ that only considers edges in $E \setminus E'$ with margin $> k$.
\end{itemize}
    
\iflong
\subparagraph{2. Assigning the Edge State (lines \ref{line:if not dominated} to \ref{line:end CyCh update}).}
\fi
If the edge is not rejected in Step 1, the algorithm will add the edge to $\FUNsym$ and determine its state.
A first preliminary state is derived from the current state of the target vertex $y$, without yet accounting for possible cycles in $\FUNsym$ -- resulting in one of the states \Efix, \EBC, or \ECBC (lines \ref{line:if not dominated} -- \ref{line:update BrCh edges}).
Later, the algorithm checks whether $(x,y)$ is contained in cycles in \FUNsym to determine whether it (and possibly other edges of margin $k$) must be updated to \ECC (lines \ref{line:begin CyCh update} -- \ref{line:end CyCh update}).

Note that the state of an edge is tentative until all edges of the same margin are processed. Only then it is clear which cycles this edge is contained in.
For this, in lines~\ref{line:begin CyCh update} -- \ref{line:end CyCh update}, the algorithm performs the \textit{cycle update check}:
it updates the edges in cycles that are closed by $(x,y)$ in \FUNsym.
To compute $C$ in \ref{line:begin CyCh update}, the algorithm performs two BFS in \FUNsym:
A forward BFS from $y$ to find all vertices $Y$ reachable from $y$ and a reverse BFS from $x$ to find all vertices $X$ reaching $x$.
Now $C$ is the set of edges $(u,v)$ with $u$ in $Y$ and $v$ in $X$.
    
\subsection{Computing the \RVput Winners in Polynomial Time} \label{subsec:algorithm-runtime}

After computing the Fused-Universe diagram \FUNsym, our algorithm returns the set of vertices that are cycle dominated or not dominated in \FUNsym as \RVput winners.
These two steps run in polynomial time in the size of the margin graph.
\begin{theorem}\label{thm:runtime-FUN}
    The runtime of the \FUN algorithm executed on a margin graph $\mgg = (A, E)$ is polynomial in \iflong the number of alternatives \fi $|A|$.
\end{theorem}
\begin{proof}
    We assume that $\mgg$ is given as adjacency lists.
    Likewise, the algorithm constructs \FUNsym using adjacency lists and stores each edge state along with the respective edge.

    Observe that $|E| \in \Theta(|A|^2)$ and that the following operations can be computed in polynomial time in $|A|$ on both graphs:
    (1) initializing and adding edges to \FUNsym;
    (2) sorting and iterating through all edges;
    (3) iterating over incoming edges of a vertex;
    (4) determining the state of a vertex;
    (5) running a breadth first search (BFS);
    (6) transposing the graph and running a reverse BFS;
    (7) updating an edge state;
    (8) iterating over all vertices.

    This covers all elemental operations of \Cref{alg:RVPUT} as well as the final pass over all vertices to determine the winner set.
    In \Cref{subsec:compute-diagram} we explain how the complex operations (namely \textit{cycle reject check} and \textit{cycle update check}) reduce to running $\mathcal{O}(|E|)$ breadth first searches.
    Thus, the \FUN algorithm overall runs in polynomial time in $|A|$.
\end{proof}

\section{The \FUN Algorithm Returns Exactly the \RVput Winners}
\label{sec:correctness}

    In the following two sections, we prove that one can correctly identify the \RVput winners using the \FUN diagram.
    \iflong
    Formally, we show the following theorem.
    \begin{theorem}\label{thm: main}
        For every preference profile \Pref and every alternative $a\in A$:
        $a\in\RVput(\Pref) \Leftrightarrow a$ is cycle dominated or not dominated in $\FUNsym(\Pref)$.
    \end{theorem}
    \fi

\subsection{Forward Direction: Fix, Branching Choice and Rejected Edges}
    \iflong
    In this section, we show the forward direction of \Cref{thm: main} which states that every \RVput winner $a$ is identified by our \FUN algorithm.
    \fi
    To that end, we \iflong first \fi establish a structural lemma relating the state of an edge in the \FUN diagram to its presence or absence in the River diagram under any tiebreaker.
    We write $\mggRiver_\tau$ as a shortcut for the River diagram $\RV(\Pref, \tau)$ of $\Pref$ under tiebreaker $\tau$.
    \begin{lemma}\label{lem:merged induction state properties}
        For every edge $(x,y)\in\mgg$ and for all tiebreakers $\tau\in\allDescOrders$ it holds that
        \begin{equation}
            state(x,y)=\textnormal{\Efix} \Rightarrow (x,y)\in\mggRiver_\tau; \label{state property fix}
        \end{equation}
        \begin{equation}
            \begin{split}
            state(x,y)\in\{\textnormal{\EBC, \ECBC}\} &\Rightarrow\\
            \exists (z,y)\in\mggRiver_\tau\colon \margin{z,y} \geq \margin{x,y}; \label{state property braching}
            \end{split}
        \end{equation}
        \begin{equation}
            (x,y)\notin\FUNsym \Rightarrow (x,y)\notin \mggRiver_\tau. 
        \label{state property rejected}
        \end{equation}
    \end{lemma}
    \begin{proof}
        Let $\lino\in\allDescOrders$ be an arbitrary tiebreaker.
        We show \Cref{state property fix,state property braching} and (\ref{state property rejected}) simultaneously via induction over the margin edges in order of \lino.
        
        \iflong The claims are trivially true for the induction base, as \FUNsym is initiated as a graph with no edges.
        \fi
        For the induction step, consider $(x,y)$ with $\margin{x,y}=k$.
        \iflong
        As induction hypothesis (IH) assume that \Cref{state property fix,state property braching,state property rejected} are true for all edges processed before $(x,y)$ according to \lino (in particular, all edges with margin greater than $k$).
        
        To prove \Cref{state property fix} and \Cref{state property braching}, we will use the following intermediate statement, which asserts that any edge in \FUNsym with a final state other than \ECC is never rejected from $\mggRiver_\lino$ by River's cycle condition.
        \fi
        \ifshort
        \begin{claim}[$\star$] \label{claim:not CyCh then no River cycle reject}
            Let $(x,y)\in\FUNsym$ with $\margin{x,y}=k$ be an edge with $state(x,y)\in\{\textnormal{\Efix, \EBC}$, $\textnormal{\ECBC}\}$. Then $(x,y)$ cannot be rejected from $\mggRiver_\lino$ by River's cycle condition.             
        \end{claim}
        \else
        \begin{claim} \label{claim:not CyCh then no River cycle reject}
            Let $(x,y)\in\FUNsym$ with $\margin{x,y}=k$ be an edge with $state(x,y)\in\{\textnormal{\Efix, \EBC}$, $\textnormal{\ECBC}\}$. Then $(x,y)$ cannot be rejected from $\mggRiver_\lino$ by River's cycle condition.             
        \end{claim}
        \begin{proof}
            Assume towards contradiction that $state(x,y)\in\{\textnormal{\Efix, \EBC}$, $\textnormal{\ECBC}\}$ but $(x,y)$ is rejected by River's cycle condition. Then in the River diagram, there exists a path $\Path{y}{x}\in\mggRiver_\lino$ with $\strength(\Path{y}{x})\geq k$, and all edges in $\Path{y}{x}$ are processed before $(x,y)$ by \lino.
            It follows from \Cref{state property rejected} of (IH) that $\Path{y}{x}\in\FUNsym$, as otherwise $\Path{y}{x}\notin\mggRiver_\lino$.

            Since $(x,y)\in\FUNsym$ and $state(x,y)\in\{\textnormal{\Efix}, \textnormal{\EBC}$, $\textnormal{\ECBC}\}$, the edge $(x,y)$ is not updated to \ECC during the cycle check of any edge (line \ref{line:begin CyCh update} of the \FUN algorithm).
            Thus, there is no cycle in \FUNsym containing $(x,y)$ at the time it is processed. In particular, there is no path from $y$ to $x$ with strength greater than $k$.
            Therefore, $\Path{y}{x}$ must contain at least one edge with margin $k$ and at least one of those margin-$k$ edges is processed after $(x,y)$ by the \FUN algorithm. Let $e'$ be the margin-$k$ edge which is processed last.

            When $e'$ is added to \FUNsym, the algorithm performs the cycle update check (line~\ref{line:begin CyCh update}).  
            Since the addition of $e'$ closes a cycle containing $(x,y)$ in $\FUNsym$ and $\margin{x,y}=\margin{e'}$, the cycle update check would update the state of $(x,y)$ to \ECC, contradicting the assumption.
            Hence, $(x,y)$ cannot be rejected from $\mggRiver_\lino$ by River’s cycle condition.
        \end{proof} \fi
        \iflong
        We now prove the induction step for all three equations. \fi

        \iflong
        \subparagraph{\Cref{state property fix}: $state(x,y)= \Efix \Rightarrow (x,y)\in\mggRiver_\tau$.}
        \fi
        By \Cref{claim:not CyCh then no River cycle reject}, the edge $(x,y)$ is not rejected from $\mggRiver_\tau$ by River's cycle condition.
        Assume towards contradiction that it is rejected from $\mggRiver_\tau$ by River's branching condition.
        Then there has to be an edge $(z,y) \in \mggRiver_\tau$ with $z \neq x$ that appears in $\tau$ before $(x,y)$.
        \iflong 
            It follows by (IH) (from the contraposition of \Cref{state property rejected}) that $(z,y) \in \FUNsym$. That means that $(z,y)$ is added either before or after $(x,y)$ to \FUNsym.
        \fi
        If~$(z,y)$ is added before $(x,y)$, then $y$ is already dominated once the algorithm processes~$(x,y)$; contradicting $state(x,y) = \Efix$.
        Conversely, if $(z,y)$ is added after, then the state of $(x,y)$ is updated to \EBC in \ref{line:update BrCh edges}; contradicting $state(x,y) = \Efix$.
        
        As a result, $(x,y)$ is not rejected by any of River's conditions, proving $(x,y)\in\mggRiver_\tau$.

        \iflong
        \subparagraph{\Cref{state property braching}: $state(x,y)\in\{\textnormal{\EBC, \ECBC}\} \Rightarrow \exists (z,y)\in\mggRiver_\tau\colon \margin{z,y} \geq k$.}
        \fi       
        By \Cref{claim:not CyCh then no River cycle reject}, the edge $(x,y)$ is not rejected from $\mggRiver_\tau$ by River's cycle condition.
        Assume $(x,y)$ is rejected by River's branching condition. Then there has to be an edge $(z,y)\in\mggRiver_\lino$ with $\margin{z,y}\geq k$ and the claim holds.
        Otherwise, $(x,y)$ is not rejected at all by River, $(x,y)\in\mggRiver_\tau$ and the claim holds since $\margin{x,y}=k$.

        \iflong
        \subparagraph{\Cref{state property rejected}: $(x,y)\notin\FUNsym \Rightarrow (x,y)\notin \mggRiver_\tau$}
        \fi
        \iflong
        Since $(x,y)\notin\FUNsym$, it was rejected from the \FUN diagram -- either by the branching reject check (line \ref{line:begin branching reject check}) or the cycle reject check (line \ref{line:begin cycle reject check}). 
        We show that in both cases, $(x,y)$ is also rejected from the River diagram.
        \fi

        First, assume $(x,y)$ was rejected by the branching reject check.
        Then $y$ is fixedly dominated and there exists an incoming edge to $y$ with margin $k'>k$ in \FUNsym. By \Cref{state property fix,state property braching} of (IH), there exists an edge $(z,y)\in\mggRiver_\lino$ with $\margin{z,y}\geq k'$. Hence, $(x,y)$ is rejected from $\mggRiver_\lino$ by River's branching condition.

        Now, assume $(x,y)$ was rejected by the cycle reject check.
        Let $U$ be the set of vertices from which there exists a path to $x$ in $\FUNsym$ with strength~$>k$ (note that $x\in U$). Then the cycle reject check condition ensures the following.
        \vspace{-0.4em}\begin{enumerate}
            \item \FUNsym contains a path from $y$ to $x$ with strength~$>k$. \label{item:path from y to x}
            \item For every $u\in U$, there exists a path $\Path{y}{u}\in\FUNsym$ from $y$ to $u$ with $\strength(\Path{y}{u})>k$ such that for every \ECC edge $(c,d)\in\Path{y}{u}$, all paths $\Path{d}{c}\in\FUNsym$ from $d$ to $c$ with $\strength(\Path{d}{c})\geq\margin{c,d}>k$ include vertex $y$. \label{item:U condition y independent}
        \end{enumerate}\vspace{-0.4em}
        We now construct a path from $y$ to $x$ in $\mggRiver_\tau$ by iteratively extending backward from $x$, and thereby show that $(x,y)$ must be rejected from $\mggRiver_\tau$ by the cycle condition.
        
            At each step, let $s$ be the current starting vertex of the path. If $s = y$, we are done.
            Otherwise, $s \in U$ and, by assumption, there exists a path from $y$ to $s$ in \FUNsym of strength~$>k$. Therefore, $s$ must be fixedly or cycle dominated. In both cases, we show that $s$ has an incoming edge $(c,s) \in \mggRiver_\tau$ with $\margin{c,s}>k$ such that $c \notin P$, allowing us to prepend $(c,s)$ to the path.
            As $y \in U$ and $U$ is finite, this backward process must eventually reach $y$, thereby completing the desired path from $y$ to $x$ in the River diagram.

            Initialize $P \coloneq  (x)$ as the partial path, and let $s \coloneq  x$ denote the current start vertex of $P$.
        \begin{description}
            \item[$s$ is fixedly dominated.]\quad By \Cref{state property fix,state property braching} of (IH), there is some edge $(c,s)\in\mggRiver_\lino$ with $\margin{c,s}>k$. Since we only follow edges from $\mggRiver_\tau$ in $P$ and $\mggRiver_\tau$ is a tree, $c\notin P$ as otherwise the edge $(c,s)$ would close a cycle in $\mggRiver_\tau$. We set $P \coloneq (c,s)\circ P$ and $s \coloneq c$.
            
            \item[$s$ is cycle dominated.]\quad
            By definition of cycle dominated vertices, $s$ has at least one incoming edge in \FUNsym and all such incoming edges are \ECC.
            Note that $c\in U$ for all incoming edges $(c,s)\in\FUNsym$ with $\margin{c,s} > k$ and, by assumption, $y$ has a path to $c$ in \FUNsym with strength~$>k$.
            As a result, there exists at least one incoming edge of $s$ with margin~$>k$ in \FUNsym, and all such incoming edges are on a path from $y$ with strength~$>k$.
            
            \iflong \smallskip\fi 
            If one such edge $(c,s)$ is also in the River diagram, then $c\notin P$ (as otherwise the edge $(c,s)$ would close a cycle in $\mggRiver_\tau$) and we set $P\coloneq (c,s)\circ P$ and $s\coloneq c$. \iflong

            \smallskip\fi Otherwise, every such edge $(c,s)\in\FUNsym$ with margin~$>k$ is rejected from $\mggRiver_\tau$.
            \begin{itemize}
                \item If $(c,s)$ is rejected by River's branching condition, \iflong then in the River diagram, \fi there is $(c',s)\in\mggRiver_\tau$ with $\margin{c',s}\geq \margin{c,s} > k$. By the contraposition of \Cref{state property rejected} \iflong of (IH)\fi, $(c',s)\in\FUNsym$ and we set $P\coloneq (c',s)\circ P$ and $s\coloneq c'$.
                
                \item If every such edge $(c,s)$ is rejected by River's cycle condition, then \iflong in the River diagram, \fi there is a path $\Path{s}{c}\in\mggRiver_\tau$ from $s$ to $c$ with strength $\geq\margin{c,s}>k$. By the contraposition of  \Cref{state property rejected} \iflong of (IH)\fi, $\Path{s}{c}\in \FUNsym$. Now, since $(c,s)$ is a \ECC edge on a path from $y$ to $s$ and $\strength(\Path{s}{c})\geq\margin{c,s}>k$, our assumption implies $y\in\Path{s}{c}$.
                
                In conclusion, there exists a path $\Path{s}{c}$ in the River diagram which includes $y$. In this case, $y$ already has an incoming edge in the River diagram $\mggRiver_\tau$ and $(x,y)$ is rejected from $\mggRiver_\tau$ by the branching condition.
            \end{itemize}
        \end{description}
        Therefore, either $(x,y)$ is rejected from $\mggRiver_\tau$ by Rivers' branching condition, or there exists a path from $y$ to $x$ in $\mggRiver_\lino$ with strength~$>\margin{x,y}$, because of which $(x,y)$ is rejected from $\mggRiver_\tau$ by River's cycle condition.

        \smallskip This proves the induction step for \Cref{state property fix,state property braching,state property rejected}, and therefore shows the claim.
    \end{proof}
    With this, we can now prove the forward direction\iflong of the correctness theorem\fi.
    \begin{theorem}[Forward Direction of Main Theorem] \label{thm:main Hinrichtung}
        For 
        every preference profile \Pref and every alternative $a \in A$:
        \ifshort
        \begin{align*}
                a \in\RVput(\Pref) \Rightarrow a \text{ is cycle/not dominated in }\FUNsym(\Pref)
        \end{align*}
        \else
        \begin{align*}
                a \in\RVput(\Pref) &\Rightarrow a \text{ is cycle or not dominated in }\FUNsym(\Pref).
        \end{align*}
        \fi
    \end{theorem}
    \begin{proof}
    We show the contraposition of the claim. For that, recall that by definition every alternative that is neither cycle dominated nor not dominated, must be fixedly dominated. The contraposition can therefore be formulated as
        $a \notin\RVput(\Pref) \Leftarrow a$ \textit{ is fixedly dominated}.
    Suppose $a$ is fixedly dominated. Then, by definition, it has either an incoming edge labeled \Efix, or at least one incoming edge labeled \EBC or \ECBC in $\FUNsym$.  
    By \Cref{lem:merged induction state properties}, there exists some incoming edge to $y$ in every universe.  
    Therefore, $a \notin \RV(\Pref, \tau)$ and $a\notin\RVput(\Pref)$.
\end{proof}
    
\subsection{Backward Direction: Cycle Edges and a Winning Certificate}
    In this section, we show that every alternative $a$ identified as an \RVput winner by the \FUN algorithm is indeed a River winner in at least one tiebreaker universe.
    To that end, we present a procedure that extracts a River diagram from \FUNsym that certifies $a$ as a winner.
    We also describe how to construct the tiebreaking order that defines this universe.

\subsubsection{Deriving a Potential Certificate for an Alternative}
\label{sec:certificate}
    Recall that, for a given preference profile $\Pref$ and tiebreaker \lino, the River Method computes a tree rooted in some alternative $a$\iflong: the \textit{River diagram} $\mggRiver(\Pref, \lino)$\fi.
    This tree certifies $a$ as the winner by containing, for every other alternative $b \neq a$, a path from $a$ to $b$.

    \iflong
    In a similar fashion, we now describe, given the Fused-Universe diagram~$\FUNsym$ and an alternative $a$, how to compute a tree $T(\FUNsym, a)$ rooted at $a$. If $a$ is an \RVput winner, this tree will serve as a certificate: it coincides with the River diagram $\mggRiver(\Pref, \lino)$ for some tiebreaker~$\lino$ under which $a$ is the unique River winner.
    If \FUNsym is clear from the context, we denote the tree $T(\FUNsym, a)$ with $\CertTreeSym$.

    \fi
    The algorithm computing $\CertTreeSym$ is a variant of Prim's algorithm for computing a Minimum Spanning Tree \iflong(MST) \cite{primShortestConnectionNetworks1957,jarnikJistemProblemuMinimalnim1930,dijkstraNoteTwoProblems1959}\fi, therefore named \CertTreeAlgo:
    \vspace{-0.2em}
    \begin{enumerate}
        \item Initialize $T = (\{a\}, \emptyset)$.
        \item \iflong As long as there are crossing edges $E^c = \{\, (u,v) \in \FUNsym \mid u \in T \land v \notin T \,\}$,
        add to $T$ an edge from $E^c$ with maximum margin.\fi
        \label{ln:prim2}
        \item Return $\CertTreeSym = T$.
    \end{enumerate}
    \vspace{-0.2em}
    An efficient implementation uses a priority queue such as Fibonacci Heaps \cite{fredmanFibonacciHeapsTheir1987} to select the edges in step~\ref{ln:prim2}.
    For a \FUN diagram with $|A|$ alternatives and $|E|$ edges, \CertTreeAlgo has a runtime in $O(|E| + |A| \log |A|)$ \cite{cormenIntroductionAlgorithms2022}.
    
    Note that since the algorithm is initialized at $a$ and only adds edges from vertices in $T$ to vertices outside $T$, the resulting $\CertTreeSym$ is a (not necessarily spanning) tree rooted in~$a$.    
    We~now~state two properties of $\CertTreeSym$. The first establishes the existence of strong paths within the tree.
    \begin{lemma}[$\star$]\label{lem:certtree-is-strong}
        If a \FUN diagram \FUNsym contains a path $\Path{a}{b}$ from $a$ to $b$, then $\CertTreeSym$ contains a path $\Path{a}{b}'$ from $a$ to $b$ with $\strength(\Path{a}{b}') \geq \strength(\Path{a}{b})$.
    \end{lemma}
    \iflong
    \begin{proof}
        Fix $\FUNsym, a, b, \Path{a}{b}$ and let $\CertTreeSym$ be the output of \CertTreeAlgo.
        Assume towards contradiction that $\CertTreeSym$ contains no such path $\Path{a}{b}'$ and distinguish two cases:
        \begin{enumerate}
            \item There is \emph{no} path from $a$ to $b$ in $\CertTreeSym$.
                Let $u$ be the last vertex on $\Path{a}{b}$ that is reachable from $a$ in $\CertTreeSym$ and let $v$ be the successor of $u$ on $\Path{a}{b}$.
                But then $(u,v)$ is a crossing edge and thus \CertTreeAlgo cannot have terminated with $\CertTreeSym$ as output due to the loop condition in step~\ref{ln:prim2}.
            \item The unique path $\Path{a}{b}'$ from $a$ to $b$ in $\CertTreeSym$ has $\strength(\Path{a}{b}') < \strength(\Path{a}{b})$.
                Let $e'$ be an edge of minimum margin in $\Path{a}{b}'$.
                Recall how \CertTreeAlgo iteratively builds $\CertTreeSym$.
                Inspect the moment, in which the algorithm picks $e'$ as edge with maximum margin from the current set of crossing edges $E^c$ and adds it to $T$.
                At this moment, $b$ is not yet reachable from $a$ in $T$.
                Thus there is a first edge $e \in E^c$ on $\Path{a}{b}$ that is not in $T$.
                However, from $\strength(\Path{a}{b}') < \strength(\Path{a}{b})$ we can infer $\strength(e') < \strength(e)$, which contradicts that $e'$ has maximum margin in $E^c$.
        \end{enumerate}
    \end{proof}
    \fi

    \iflong
    \begin{figure*}
        \centering
        \includegraphics[width=0.8\linewidth]{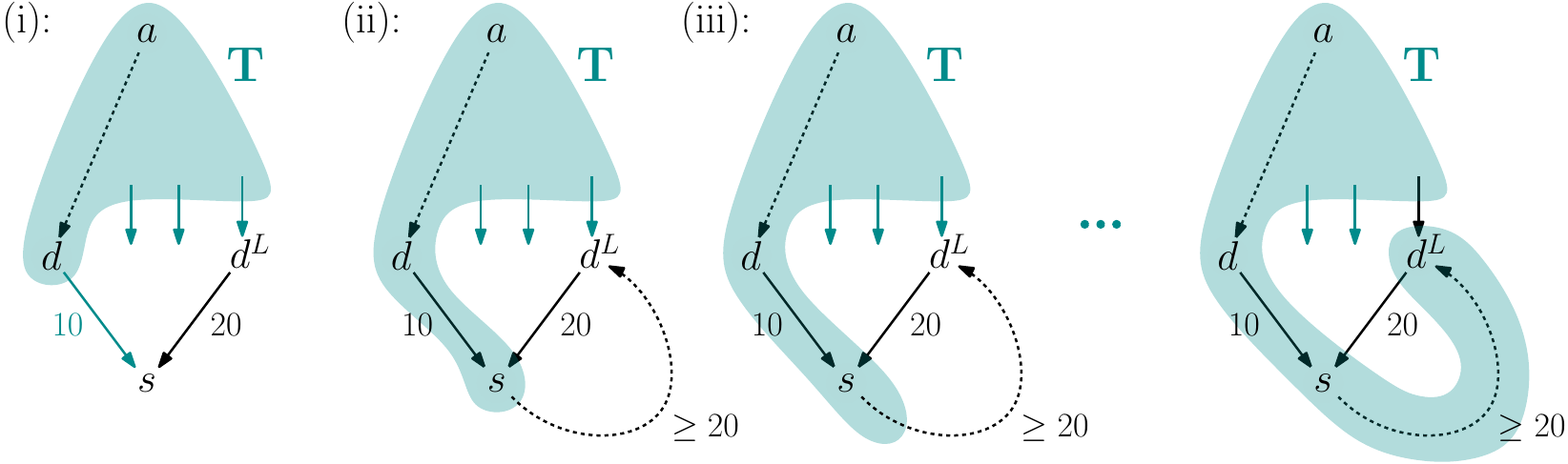}
        \caption{
        Exemplary illustration for the proof of \Cref{lem:certtree-has-path}.
        Dotted arrows represent paths, and the shaded region indicates the current tree~$T$.
        All solid arrows are edges in the Fused-Universe diagram $\FUNsym$.
        From left to right: 
        (i) before \CertTreeAlgo picks $(d,s)$ from the current set of crossing edges, with $d^L \notin T$;
        (ii) after $(d,s)$ is added to $T$ and a path $\Path{s}{d^L}$ exists in $\FUNsym$;
        (iii) when \CertTreeAlgo explores $d^L$ via a path from $s$.}
        \label{fig:illustration certificate cycle choice}
    \end{figure*}
    \fi
    \ifshort
    The next explains how \CertTreeSym may include an edge despite a higher-margin edge to the same alternative in \FUNsym. 
    \fi 
    \iflong 
    The next lemma explains how an edge can belong to the certificate tree even when a higher-margin edge to the same alternative exists. 
    See \Cref{fig:illustration certificate cycle choice} for an illustration. Specifically, if an edge $(d, s)\in\FUNsym$ is chosen for the tree $\CertTreeSym$ despite the existence of a higher-margin edge $(d^L, s)\in \FUNsym$, then $(d^L, s)$ must be an \ECC edge. This implies the existence of a strong enough path from $s$ to $d^L$ in \FUNsym, which can be used in place of the direct edge $(d^L, s)$.\fi
    \begin{lemma}[$\star$]\label{lem:certtree-has-path}
        Let \FUNsym be a \FUN diagram with margins $m$.
        Let $a$ be an alternative and let $s, d, d^L$ be different alternatives reachable from $a$ in \FUNsym.
        If \FUNsym contains edges $(d, s), (d^L, s)$ with $\margin{d, s} < \margin{d^L, s}$ and $\CertTreeSym$ contains $(d, s)$, then $\CertTreeSym$ contains a path $\Path{s}{d^L}$ from $s$ to $d^L$ with $\strength(\Path{s}{d^L}) \geq \margin{d^L, s}$.
    \end{lemma}
    \iflong
    \begin{proof}
        Fix $\FUNsym, a, s, d, d^L$ with $(d,s),(d^L,s)\in\FUNsym$, $\margin{d, s} < \margin{d^L, s}$ and $(d, s) \in \CertTreeSym$.
    
        Inspect the moment, in which \CertTreeAlgo picks $(d, s)$ as edge with maximum margin from the current set of crossing edges $E^c$ and adds it to $T$.
        At that moment, $d^L \notin T$; otherwise $(d^L, s)$ would be a crossing edge with higher margin. For an illustration, see \Cref{fig:illustration certificate cycle choice} (i).
    
        Since $(d, s), (d^L, s) \in \FUNsym$, it follows that $(d^L, s)$ must be a \ECC edge:
        otherwise $s$ would be fixedly dominated by an edge with margin at least $\margin{d^L, s}$ and no edge with margin lower than $\margin{d^L, s}$ would have been added to \FUNsym.
        By definition of \ECC edges there exists a path $\Path{s}{d^L}'$ from $s$ to $d^L$ in \FUNsym with $\strength(\Path{s}{d^L}') \geq \margin{d^L, s}$.
        (See \Cref{fig:illustration certificate cycle choice} (ii).)
        
        From $\margin{d^L, s} > \margin{d, s}$ follows that all edges in $\Path{s}{d^L}'$ have margin greater than that of any other edge in $E^c$.
        Therefore, \CertTreeAlgo explores $d^L$ via a path $\Path{s}{d^L}$ starting in $s$ with $\strength(\Path{s}{d^L}) \geq \strength(\Path{s}{d^L}')$ before considering any other edges in $E^c$.
        This is illustrated in \Cref{fig:illustration certificate cycle choice} (iii).
    \end{proof}   
    \fi

    From the certificate tree $\CertTreeSym$, we derive a tiebreaker $\lino^a$ that prioritizes the edges in $\CertTreeSym$ over any other edges with the same margin that are not in $\CertTreeSym$.
    We will show in \Cref{thm:main Rückrichtung} that whenever $a$ is not dominated or cycle dominated in \FUNsym, running River with this tiebreaker will result in a River diagram $\mggRiver(\Pref, \lino^a)$ which exactly matches $\CertTreeSym$.

    \begin{definition}\label{def:certtree-lino}
        Let $\Pref$ be a preference profile and $\mgg(\Pref)$ the corresponding margin graph.
        A \emph{\FUNlinoSymNAME} $\FUNlinoSym$ for an alternative $a$ with certificate tree $\CertTreeSym$ is an ordering of all edges in $\mgg(\Pref)$ that fulfills the following two properties:
    
        \begin{enumerate}
            \item Edges are ordered by descending margin.
            \item Among edges with identical margin, those that are in $\CertTreeSym$ come before those not in $\CertTreeSym$.
        \end{enumerate}
    \end{definition}

\subsubsection{Confirming \RVput-Winners Using the Certificate Tree}
    Note that $\CertTreeSym$ is not necessarily a spanning tree, as some alternatives may not be reachable from $a$ in \FUNsym.
    However, if $a$ is cycle dominated or not dominated, then the certificate tree $\CertTreeSym$ is a spanning tree over all alternatives:
    \begin{lemma}[$\star$]\label{lem:cycle_or_not_dom-Ta_is_spanning}
        If an alternative $a$ is cycle dominated or not dominated in $\FUNsym$ then the certificate tree $T^a$ is a spanning tree of \mgg rooted in $a$.
    \end{lemma}
    \iflong
    \begin{proof} For the definitions of cycle and not dominated, refer to \Cref{tab:FUN_alg-edge-states}. We distinguish between $a$ being not dominated and $a$ being cycle dominated, and show that $a$ can reach every other alternative in \FUNsym in both cases.
        \subparagraph{Case 1: $a$ is not dominated in \FUNsym.}
        This means $a$ has no incoming edge in \FUNsym.
        Consider any vertex $x \in A\setminus\{a\}$. We walk backward from $x$ using only incoming edges labeled \Efix, \EBC, or \ECBC until one of the following happens:
        \begin{enumerate}
            \item[(i)] We reach $a$: then clearly $a$ can reach $x$ via the reverse of this walk.
            \item[(ii)] We visit a vertex twice: then we have found a cycle in \FUNsym composed solely of \Efix, \EBC, or \ECBC edges. But this contradicts the \FUN algorithm’s handling of cycles: any such cycle must contain (i) a unique weakest edge, which would have been rejected by the cycle reject check; or (ii) multiple weakest edges, all of which would have been updated to \ECC in the cycle update check.
            \item[(iii)] We get stuck at a vertex $x'$ where all incoming edges in \FUNsym are labeled \ECC. Then $x'$ is cycle dominated, and we distinguish two subcases:
            \begin{itemize}
                \item If $(a, x') \in \mgg$, then $(a,x')$ was \emph{not} rejected by the branching condition (since $x'$ is not fixedly dominated), and not rejected by the cycle condition (since $a$ has no incoming edge in \FUNsym). Hence, $(a,x') \in \FUNsym$.
                \item If $(x',a) \in \mgg$, then since $a$ is not dominated, $(x',a)\notin\FUNsym$ and it was rejected by the cycle reject check. That implies that there exists a path from $a$ to $x'$ in \FUNsym of strength at least $\margin{x',a}$.
            \end{itemize}
             In either subcase, $a$ can reach $x'$ and thereby $x$.
        \end{enumerate}

        \subparagraph{Case 2: $a$ is cycle dominated in \FUNsym.}
        Then $a$ has at least one incoming edge in \FUNsym and all such edges are \ECC. Consider any vertex $x \in A\setminus\{a\}$. As in Case 1, we walk backward from $x$ using only incoming \Efix, \EBC, or \ECBC edges until one of the following happens:
            \begin{enumerate}
                \item[(i)] We reach $a$: then clearly $a$ can reach $x$ via the reverse of this walk.
                \item[(ii)] We visit a vertex twice: then we have found a cycle in \FUNsym composed solely of \Efix, \EBC, or \ECBC edges. As in Case 1, this contradicts the \FUN algorithm’s handling of cycles (see Case 1 (ii)).
                \item[(iii)] We get stuck at a vertex $x'$ where all incoming edges in \FUNsym are labeled \ECC. Then $x'$ is cycle dominated, and we distinguish two subcases:
                \begin{description}
                    \item[$(a, x') \in \mgg$.]\quad If $(a,x')\in\FUNsym$, then $a$ can reach $x'$. 
                    
                    Otherwise,~$(a,x')\notin\FUNsym$. Since $x'$ is not fixedly dominated, $(a,x')$ was not rejected from \FUNsym by the branching reject check, and thus was rejected by the cycle reject check.
                    This implies a path from $x'$ to $a$ in \FUNsym of strength~$>\margin{a,x'}$, and for every \ECC edge $(c,d)$ on such a path, the cycle reject condition ensures that all paths from $d$ to $c$ with margin $\geq \margin{c,d}$ go through $x'$.
                    In particular, the edge $(c,a)$ at the end of such a path must satisfy this condition: since $(c,a)$ is \ECC (as $a$ is cycle dominated), there exists a path from $a$ to $c$ with strength $\geq \margin{c,a}$ and by the condition above, this path must pass through $x'$. Hence, $a$ can reach $x'$.
                    \item[$(x',a) \in \mgg$.]\quad If $(x',a)\in\FUNsym$, then it must be \ECC since $a$ is cycle dominated. This implies that there exists a path from $a$ to $x'$ in \FUNsym. 
                    
                    Otherwise, $(x',a)\notin\FUNsym$. Since $a$ is not fixedly dominated, $(x',a)$ was not rejected from \FUNsym by the branching reject check, and thus was rejected by the cycle reject check. This implies that there is a path from $a$ to $x'$ in \FUNsym. 
                \end{description}
                 In either subcase, $a$ can reach $x'$ and thereby $x$.
            \end{enumerate}
        In both Case 1 and Case 2, $a$ has a directed path in \FUNsym to every other alternative. \Cref{lem:certtree-is-strong} implies that \CertTreeSym is a spanning tree of $\FUNsym$ rooted in $a$.
    \end{proof}
    \fi
    \iflong With this, we are ready to prove the backward direction of \Cref{thm: main}.\fi
    \ifshort
    The full proof can be found in the full version of the paper.
    In it, we show that the certificate tree \CertTreeSym for $a$ is exactly the River diagram under the \FUNlinoSymNAME $\FUNlinoSym$. Since \CertTreeSym is rooted in $a$, it follows that $a$ is the unique River winner in that universe.
    The proof is by induction over the edges, following $\FUNlinoSym$, and relies on structural properties of the certificate tree established in \Cref{lem:certtree-is-strong,lem:certtree-has-path}.
    \fi
    \begin{theorem}[\ifshort($\star$)\fi Backward Direction of Main Theorem] \label{thm:main Rückrichtung}
        For every preference profile \Pref and every alternative $a \in A$:
        \ifshort
        \begin{align*}
                a \in\RVput(\Pref) \Leftarrow a \text{ is cycle/not dominated in }\FUNsym(\Pref)
        \end{align*}
        \else
        \begin{align*}
                a \in\RVput(\Pref) &\Leftarrow a \text{ is cycle or not dominated in }\FUNsym(\Pref).
        \end{align*}
        \fi
    \end{theorem}

\newcommand{\proofRVdiagram}{\ensuremath{\mggRiver_{\FUNlinoSym}}\xspace}

\newcommand{\Lvertex}{d^L}
\newcommand{\Svertex}{d}
\newcommand{\Invertex}{s}

\newcommand{\zvertex}{x^L}

\newcommand{\cvertex}{\ensuremath{u}\xspace}
\newcommand{\fvertex}{\ensuremath{u'}\xspace}
\newcommand{\dvertex}{\ensuremath{v}\xspace}

\newcommand{\edgeIH}{e'}

\iflong
\begin{proof}
    Let \Pref be a preference profile and let $a\in A$ be an alternative. 
    Recall the definition of \RVput which states $a\in\RVput(\Pref) \Leftrightarrow \exists \tau\in\allDescOrders\colon \RV(\Pref, \tau) = \{a\}$. Thus, we have to show: 
    \begin{align*}
        \exists {\lino} \in \allDescOrders : \RV(\Pref, \tau)=\sset{a} \Leftarrow &a \text{ is cycle dominated or}\\ &\text{ not dominated in } \FUNsym(\Pref).
    \end{align*}
    Assume $a$ is either cycle dominated or not dominated in $\FUNsym$. Then by \Cref{lem:cycle_or_not_dom-Ta_is_spanning}, the certificate tree $T^a$ of $a$ is a spanning tree.
    Let $\FUNlinoSym$ be the corresponding \FUNlinoSymNAME.
    We prove that $\RV(\Pref,\FUNlinoSym)=\{a\}$ by establishing the stronger result that the River diagram under \FUNlinoSym is exactly the certificate tree rooted in $a$, \ie $\mggRiver_{\FUNlinoSym} = \CertTreeSym$. 

    Since \CertTreeSym and \proofRVdiagram are trees on the same vertex set, it suffices to show that one edge set is subset of the other. We show $\CertTreeSym\subseteq\proofRVdiagram$ by induction over the edges of $\FUNsym$ in the order $\FUNlinoSym = (e_0, e_1, \ldots, e_\ell)$.
    For the induction base consider the edge $e_0$ and assume $e_0\in\CertTreeSym$.
    Since $e_0$ is the first edge processed by River, $e_0$ cannot be rejected by $(Br)$ or $(Cy)$, and consequently $e_0\in\proofRVdiagram$.

    Now, let $0 < i < \ell$ and assume as induction hypothesis (IH) for $0 \leq j < i$ that $e_j \in \CertTreeSym$ implies $e_j \in \proofRVdiagram$.
    We use this to show that $e_i\in\CertTreeSym$ implies $e_i\in\proofRVdiagram$.
    Denote the endpoints of the edge as $(x, y) = e_i$ and assume towards contradiction that $(x,y)\in\CertTreeSym$ but $(x,y)\notin\proofRVdiagram$.
    This implies that $(x,y)$ was rejected from \proofRVdiagram by River's branching condition or cycle condition.
    We distinguish accordingly:

    \subparagraph{Case 1: $(x,y)$ is rejected by the branching condition.}
    Consequently, there exists an edge $(\zvertex,y)\in\proofRVdiagram$ with $\margin{x,y}\leq \margin{\zvertex,y}$.
    Since \CertTreeSym is a tree with at most one incoming edge per vertex and $(x,y)\in\CertTreeSym$, we conclude $(\zvertex,y)\notin\CertTreeSym$.

    If~$\margin{\zvertex,y}=\margin{x,y}$, then by definition of \FUNlinoSym, $(x,y)$ is processed by River before $(\zvertex,y)$. But then $(x,y)$ cannot be rejected from \proofRVdiagram by $(Br)$ because of $(\zvertex,y)$, since $(\zvertex,y)$ has not even been processed yet.

    Therefore, we can assume $\margin{x,y}<\margin{\zvertex,y}$.
    Now, observe that $(x,y)\in\FUNsym$ and $(\zvertex,y)\in\FUNsym$. It follows from \Cref{lem:certtree-has-path} with $\Svertex=x$, $\Lvertex=\zvertex$, and $\Invertex=y$ that $\CertTreeSym$ includes a path from $y$ to $\zvertex$ with strength larger $\margin{x,y}$.
    By (IH) $\proofRVdiagram$ includes the same path and thus cannot include $(\zvertex,y)$ -- a contradiction.

    \subparagraph{Case 2: $(x,y)$ is rejected by the cycle condition.}
    Consequently, there is a path $\Path{y}{x}\in\proofRVdiagram$ with $\margin{x,y}\leq \strength(\Path{y}{x})$.
    Since \CertTreeSym is acyclic, \Path{y}{x} cannot be fully contained in \CertTreeSym, \ie there must be at least one edge $(\cvertex,\dvertex)\in\Path{y}{x}$ with $(\cvertex,\dvertex)\notin\CertTreeSym$ and $\margin{x,y}\leq\margin{\cvertex,\dvertex}$. We show that this implies $(\cvertex,\dvertex)\notin\proofRVdiagram$, which contradicts $\Path{y}{x}\in\proofRVdiagram$.

    If $\margin{x,y}=\margin{\cvertex,\dvertex}$, then by definition of \FUNlinoSym, $(x,y)$ is processed by River before $(\cvertex,\dvertex)$.
    But then $(x,y)$ cannot be rejected from \proofRVdiagram by $(Cy)$ because of $P_{yx}$, since $(\cvertex,\dvertex)$ has not even been processed yet and consequently $P_{yx}$ is not yet in \proofRVdiagram.
    
    Therefore, we can assume $\margin{x,y}<\margin{\cvertex,\dvertex}$.
    Now, from $(\cvertex,\dvertex)\notin\CertTreeSym$ follows that either (1) there is some other edge $(\fvertex,\dvertex)\in\CertTreeSym$ which is chosen instead of $(\cvertex,\dvertex)$, or (2) there is no edge towards $\dvertex$ in \CertTreeSym.
    We distinguish accordingly:
    \begin{description}
        \item[Case 2.1]\quad 
            Assume there is some other edge $(\fvertex, \dvertex) \in \CertTreeSym$ which is chosen instead of $(\cvertex, \dvertex)$.
            Observe that $(\fvertex, \dvertex) \in \FUNsym$ and $(\cvertex, \dvertex) \in \FUNsym$.
           If $\margin{x,y}<\margin{\fvertex,\dvertex}$, then $(\fvertex,\dvertex)\in\proofRVdiagram$ follows from (IH).
           In that case, $(\cvertex, \dvertex)$ would be rejected by $(Br)$ from $\proofRVdiagram$, \ie $(\cvertex, \dvertex) \notin \proofRVdiagram$.
           If~$\margin{\fvertex, \dvertex} \leq \margin{x,y}$, then $\margin{\fvertex,\dvertex}< \margin{\cvertex, \dvertex}$ and it follows from \Cref{lem:certtree-has-path} with $\Svertex=\fvertex$, $\Lvertex=\cvertex$, and $\Invertex=\dvertex$ that $\CertTreeSym$ includes a path from $\dvertex$ to $\cvertex$ with strength larger $\margin{\cvertex,\dvertex} > \margin{x,y}$.
           By (IH), $\proofRVdiagram$ includes the same path and thus cannot include $(\cvertex,\dvertex)$.
        \item[Case 2.2]\quad
            Assume there is no edge towards $\dvertex$ in \CertTreeSym.
            In that case, $\dvertex$ must be the root $a$, and we have to show $(\cvertex,a)\notin\proofRVdiagram$.
            As $a$ has only incoming \ECC edges in \FUNsym, $(\cvertex,a)$ has to be a \ECC edge.
            By the definition of \ECC there exists a path \Path{a}{\cvertex} from $a$ to $\cvertex$ in \FUNsym with $\strength(\Path{a}{\cvertex}) \geq \margin{\cvertex,a}$.
            It follows from \Cref{lem:certtree-is-strong} that $\CertTreeSym$ contains a path $\Path{a}{\cvertex}'$ from $a$ to $\cvertex$ with $\strength(\Path{a}{\cvertex}') \geq \strength(\Path{a}{\cvertex})$

            As $\strength(\Path{a}{\cvertex}) > \margin{x,y}$ (IH) implies that $\Path{a}{\cvertex}' \in \proofRVdiagram$ and $(\cvertex,a)$ would be rejected from \proofRVdiagram by $(Cy)$, \ie $(\cvertex,\dvertex)=(\cvertex,a)\notin\proofRVdiagram$.
    \end{description}
\end{proof}
\fi

\iflong
    \section{Experiments}
    \begin{figure*}
        \centering
        \includegraphics[width=0.9\linewidth]{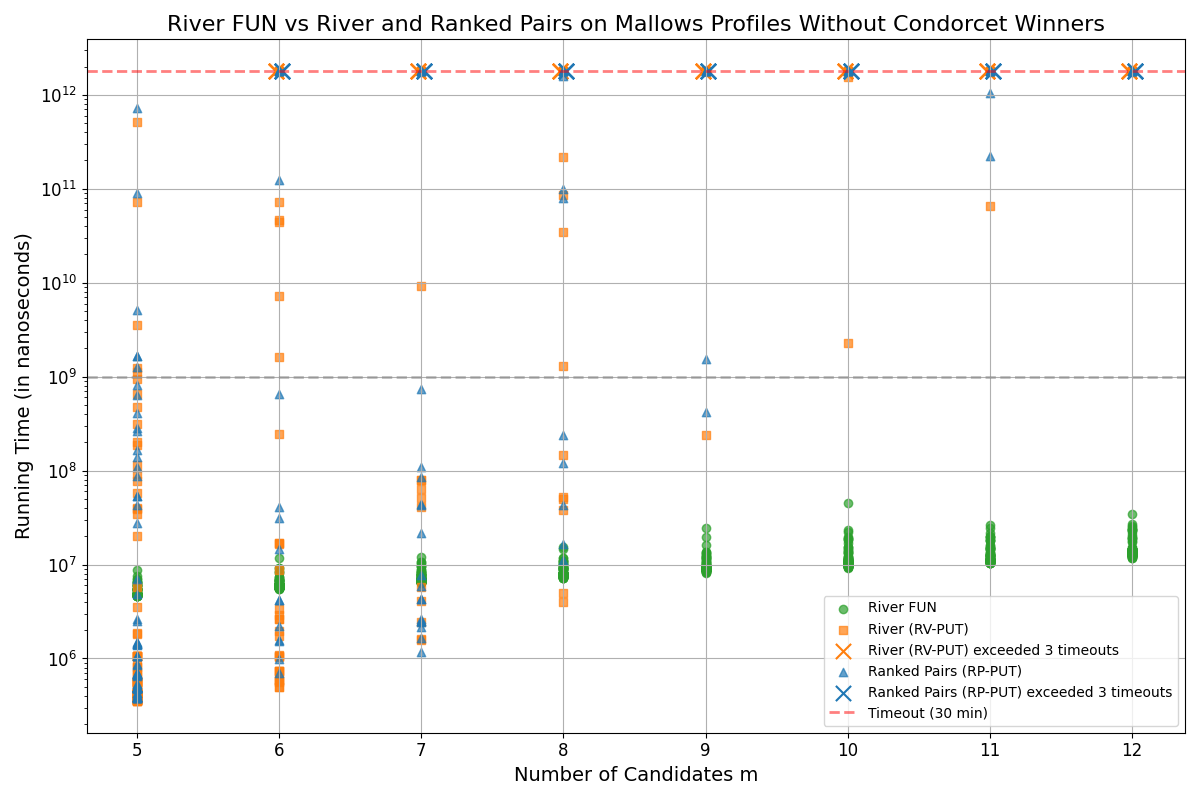}
        \caption{
        A scatterplot of all recorded running times of the \FUN algorithm versus \RVput and \RPput as implemented in pref-voting by \cite{mattei_preflib_2013} over  synthetic election with different numbers of alternatives and voters on a logarithmic scale. 
        Note that after 3 timeouts, no more running times where recorded for that $m$.
        }
        \label{fig:Fun_vs_RVandRPPUT}
    \end{figure*}
    \begin{figure*}
        \centering
        \includegraphics[width=\linewidth]{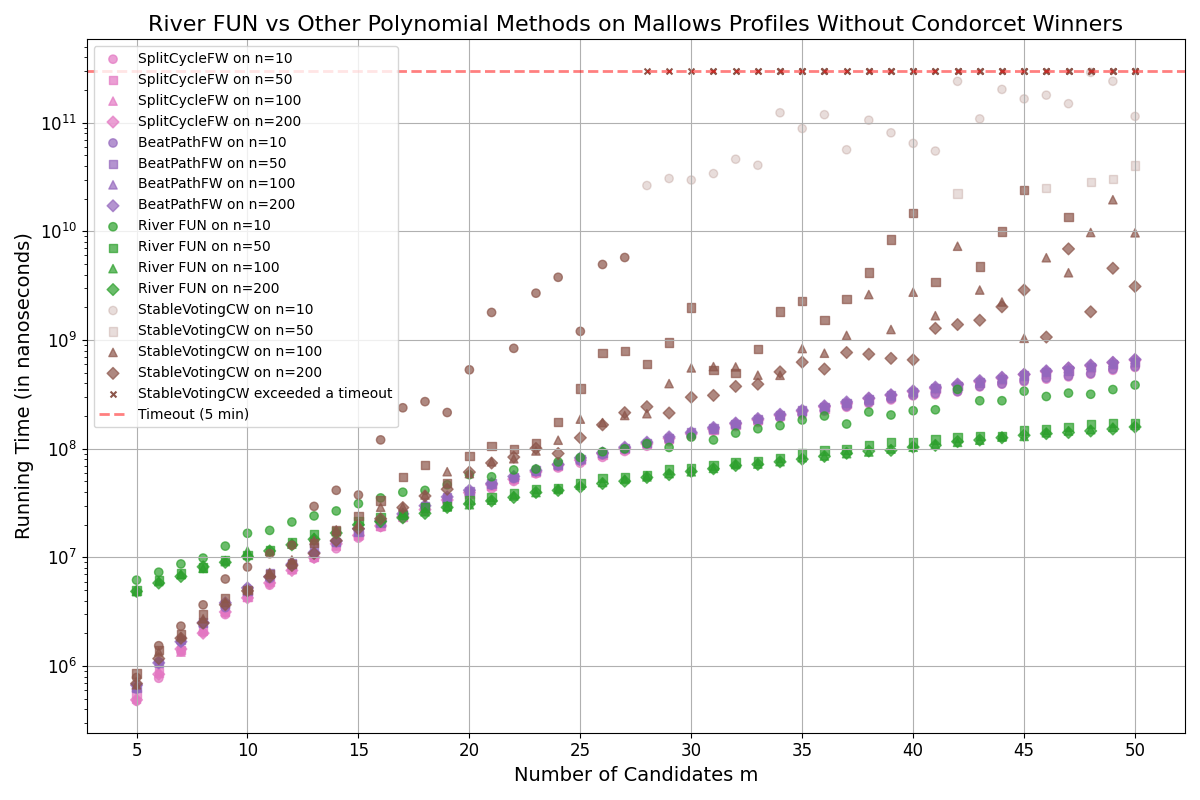}
        \caption{
        Average running time of the \FUN algorithm versus Stable Voting, Beat Path, and Split Cycle as implemented in pref-voting by \cite{mattei_preflib_2013} over single runs per 20 synthetic elections with different numbers of alternatives and voters on a logarithmic scale.
        }
        \label{fig:Fun_vs_OtherP}
    \end{figure*}

         %
    To assess the practical performance of the \FUN algorithm, we conducted preliminary experiments comparing its running time to that of other voting methods. In particular, we evaluated River \FUN against River with brute-force PUT (\RVput), Ranked Pairs with brute-force PUT (\RPput), as well as the polynomial-time methods Split Cycle, Stable Voting, and Beat Path, using implementations from the pref\_voting Python library~\cite{HollidayPacuit2025}.

    \subsection{Setup}
    We followed the recommendations by \cite{szufa_drawing_2025} and generated synthetic elections using the normalized Mallows model with dispersion parameter $\text{norm-}\phi=0.7$, which simulates elections that resemble real-world data.
    We focused only on elections without a Condorcet winner, since all considered methods are Condorcet consistent and first check for Condorcet winners, therefore taking the same amount of time to return the winner. We generated elections for all combinations of:\begin{itemize}
        \item number of alternatives $m \in \sset{5,6,\dots,50}$, and
        \item number of voters $n \in \sset{10, 50, 100, 200}$.
    \end{itemize}
    For each $(m,n)$ pair, we generated elections until we obtained 20 elections without a Condorcet winner or until we discarded 50,000 samples.
    For the parameter combinations $m=5$ with $n=50$, for $m \in \sset{5,6,7,8,9}$ with $n=100$, as well as for $m \in \sset{5, \dots, 15}$ with $n=200$ where we could not generate 20 suitable elections, we excluded them from the results.

    All voting methods were run sequentially in randomized order across all eligible elections. To keep computing time within a feasible timeframe, we applied the following timeouts: \begin{itemize}
        \item 5 minutes for polynomial-time methods (Split Cycle, Stable Voting, Beat Path and River \FUN), and
        \item 30 minutes for \RVput and \RPput.
    \end{itemize}
    If \RVput or \RPput exceeded the timeout $3$ times for one $m$ they where not run again for that $m$. Since pilot experiments indicated that both methods would exceed the timeout at small election sizes, we limited their experiments to $m \in \sset{5, 6, \dots, 12}$.
    The polynomial-time methods where not rerun for $(m,n)$ once if they had exceeded their timeout.

    The experiments were run on a $2.30$ GHz core with 64GB RAM under Ubuntu 24.04.2 LTS, using CPython 3.12.3 with pref-voting 1.16.21, multiprocess 0.70.18, and networkx 3.5.

    \subsection{Goals}
    The goal of our experiments was twofold:\begin{description}
        \item[Scalability] Determine whether the \FUN algorithm makes computing \RVput feasible on instances where brute-force methods fail.
        \item[Performance] Compare \FUN’s running time to that of other common polynomial-time voting methods.
    \end{description}
    
    \subsection{Results}
    Our findings are illustrated in \Cref{fig:Fun_vs_RVandRPPUT} and \Cref{fig:Fun_vs_OtherP}.
    \begin{description}
        \item[River \FUN vs \RVput and \RPput] Our results confirm that the  \FUN algorithm makes River with Parallel-Universe Tiebreaking computationally feasible even for moderately sized elections. As shown in \Cref{fig:Fun_vs_RVandRPPUT}, both \RVput and \RPput exceeded the 30-minute timeout on more than three instances for $m \geq 7$. In contrast, the \FUN algorithm consistently computed the River outcome in under 0.1 seconds—on the same instances.
        
        \item[River \FUN vs. Other Polynomial-Time Methods] 
        \FUN’s performance is comparable to that of Split Cycle, Beat Path, and Stable Voting (\Cref{fig:Fun_vs_OtherP}). The implementations of Split Cycle and Beat Path in \cite{mattei_preflib_2013} both use the Floyd–Warshall algorithm, which explains their nearly identical performance. While \FUN is slightly slower than the other methods for $m \leq 16$, its performance surpasses Stable Voting from $m \geq 17$ and surpasses/ is on par with Beat Path and Split Cycle from $m \geq 20$. Note that for $n =10$, its running time closely matches that of Beat Path and Split Cycle.

        We also observed that \FUN’s performance varies with the number of voters: when $n$ is small relative to $m$, the computation time increases. We conjecture this is due to a higher number of nearly tied alternatives, causing more edges to be retained in the \FUN diagram.

        Finally, we observed that Stable Voting exhibits greater variance in its running time compared to the other methods, which show smooth scaling. This is likely due to Stable Voting’s recursive definition, which can require many repetitions of winner computation depending on the profile structure.
    \end{description}

\fi

\section{Conclusion}
\label{sec:conclusion}
We presented a polynomial-time algorithm for computing the PUT (Parallel-Universe Tiebreaking) variant of River.
Our \FUN algorithm exploits River's branching condition to maintain a compact diagram capturing all possible River outcomes across tiebreakers.
This highlights the computational benefit of River's simple decision process, especially in contrast to Ranked Pairs, whose PUT variant is \NP-hard.

\iflong
While the \FUN diagram compactly encodes all possible River outcomes, it may include edges that do not appear in any River diagram
Our algorithm only removes edges which are rejected by River in all universes for the same reason. Thus it does not remove edges rejected sometimes by branching and sometimes by cycle condition.
While our algorithm is sufficient for correctness, further work could refine this to compute the exact union of all River diagrams.
\fi

As \RVput defines a new social choice function, an axiomatic study could be of interest. In particular in comparison to PUT-Ranked Pairs and River in terms of winning set containment and axioms such as clone independence.


Preliminary experiments indicate that \FUN scales to instances where brute-force PUT fails and is competitive with well-regarded polynomial-time rules. A broader empirical study on synthetic and real data, including comparison to the optimized \RPput implementation by Wang et al.\footnote{We thank an anonymous AAAI~2026 reviewer for this pointer.}, is promising future work.



\section*{Acknowledgments}
The project on which this report is based was funded by the Federal Ministry of Research, Technology and Space under the funding code “KI-Servicezentrum Berlin-Brandenburg” 16IS22092. Responsibility for the content of this publication remains with the author (Michelle Döring).


\bibliography{references,zot-references}



\end{document}